\documentclass[
    12pt,
    a4paper
    ,pop
    ,oneside
    ,nobibnotes
    ,floatfix
    ,showpacs
]{revtex4}

\usepackage{ifpdf}
\ifpdf
    \usepackage{cmap}
\fi

\usepackage{graphicx}\graphicspath{{Graph/}{../Graph/}}

\usepackage[utf8]{inputenc}
\usepackage{amsmath,amssymb,bm}

\ifpdf
    \usepackage[pdftex,unicode,colorlinks,pagebackref
        ,pdfpagemode=FullScreen,
        ,hypertexnames=false
        ,bookmarksnumbered
        ,pdfpagemode=UseOutlines
        ,pdfdisplaydoctitle=true
        ,pdfstartview=FitH
    ]{hyperref}
    \hypersetup{
        pdftitle={Electron cyclotron resonance near the axis of the gas-dynamic trap}
        ,pdfauthor={D. S. Bagulov, I. A. Kotelnikov}
        ,pdfsubject={Draft, \today}
        ,pdfkeywords={electron cyclotron resonance, gas-dynamic trap}
    }
\fi

    \newcommand{\re}{\mathrm{Re}}
    \newcommand{\im}{\mathrm{Im}}

    \newcommand{\parder}[2]{\frac{\partial #1}{\partial #2}}

    \DeclareMathOperator{\dif}{d\!}
    \newcommand{\der}[2]{\frac{\dif{#1}}{\dif{#2}}}

    \renewcommand{\vec}[1]{\mathbf{#1}}

    \usepackage{xcolor}
    \definecolor{darkblue}{cmyk}{1.00, 0.50, 0.00, 0.40}

\begin{document}

\title{Electron cyclotron resonance near the axis of the gas-dynamic trap}

\author{D.\,S.~Bagulov}
    \email{bagulov-denis@yandex.ru}
    \affiliation{Novosibirsk State University}

\author{I.\,A.~Kotelnikov}
    \email{I.A.Kotelnikov@inp.nsk.su}
    \affiliation{Budker Institute of Nuclear Physics SB RAS}
    \affiliation{Novosibirsk State University}

\begin{abstract}
    Propagation of an extraordinary electromagnetic wave in the vicinity of electron cyclotron resonance surface in an open linear trap is studied analytically, taking into account inhomogeneity  of the magnetic field in paraxial approximation. Ray trajectories are derived from a reduced dispersion equation that makes it possible to avoid the difficulty associated with a transition from large propagation angles to the case of strictly longitudinal propagation. Our approach is based on the theory, originally developed by the Zvonkov and Timofeev \cite{ZvonkovTimofeev1988SovJPlasmaPhys_14_743}, who used the paraxial approximation for the magnetic field strength, but did not consider the slope of the magnetic field lines, which led to considerable error, as has been recently noted by Gospodchikov and Smolyakova \cite{Gospodchikov+2011PPR_37_768}. We have found ray trajectories in analytic form and demonstrated that the inhomogeneity of both the magnetic field strength and the field direction can qualitatively change the picture of wave propagation and significantly affect the efficiency of electron cyclotron heating of a plasma in a linear magnetic trap. Analysis of the ray trajectories has revealed a criterion for the resonance point on the axis of the trap to be an attractor for the ray trajectories. It is also shown that a family of ray trajectories can still reach the resonance point on the axis if the latter generally repels the ray trajectories.

    As an example, results of general theory are applied to the electron cyclotron resonance heating experiment which is under preparation on the Gas Dynamic Trap in the Budker Institute of Nuclear Physics \cite{Shalashov+2012PoP_19_052503}.

\end{abstract}



\maketitle

\section{Introduction}
\label{s1}

The behavior of electromagnetic waves in plasmas in the range of the electron cyclotron  frequency remains the subject of research for a long time. The main results of these studies are well known---see, for example, \cite{Ginzburg1970, Stix1992, Geller1996, Timofeev2009(en)}, and the interest in the problem is supported at high level due to important applications such as high-frequency heating and plasma diagnostics in magnetic traps.

The theory of propagation of electromagnetic waves in a plasma with a one-dimensional inhomogeneity is developed in detail.  In particular, in a plane-layered system, spatial resonance is a singular point of the wave equation  \cite{Timofeev2009(en)}. In terms of ray trajectories, it means that the movement of a wave packet slows down along the inhomogeneity as the packet approaches the point (plane) of spatial resonance. This leads to the accumulation of wave energy in the vicinity of the resonance point and the appearance of singularities in the distribution of electromagnetic fields.

In the case of two-dimensional inhomogeneity, determining the position of the resonance surface is more complicated. According to the pioneering paper \cite{PiliaFedorov1992}, it can be done by analyzing the trajectories of wave packets (ray traces).

Due to the short wavelength in the range of frequencies of electron cyclotron resonance, the wave propagation in laboratory magnetic traps is described with reasonable  accuracy by the approximation of geometric optics, and the investigation of the ray trajectories can provide both qualitative and quantitative assessment of the possibility of plasma heating in the trap.

In this paper, propagation of an extraordinary electromagnetic wave in the vicinity of electron cyclotron resonance surface in an open linear trap is studied analytically with taking into account the heterogeneity of the magnetic field in paraxial approximation. Wave trajectories are described by solving a reduced dispersion relation that makes it possible to avoid difficulties associated with a transition from large propagation angles to the case of strictly longitudinal propagation. Our approach is based on the theory, initially developed by Zvonkov and Timofeev \cite{ZvonkovTimofeev1988SovJPlasmaPhys_14_743}, who used paraxial approximation for the magnetic field strength. Unfortunately, they ignored the inclination the magnetic field lines, which resulted in significant error as was recently noted by Gospodchikov and Smolyakova \cite{Gospodchikov+2011PPR_37_768}. We refine the theory of Ref.~\cite{ZvonkovTimofeev1988SovJPlasmaPhys_14_743} in order to take into account the effect specified in Ref.~\cite{Gospodchikov+2011PPR_37_768}. The refined theory makes it possible to find ray trajectories in analytic form.

As an example, we apply the results of general theory to the electron cyclotron resonance heating (ECRH) experiment at the Gas Dynamic Trap (GDT) which is under preparation in the Budker Institute in Novosibirsk \cite{Shalashov+2012PoP_19_052503}. The key physical issue of the GDT magnetic field topology is that conventional ECRH geometries are not accessible. The solution, proposed and numerically investigated in \cite{Shalashov+2012PoP_19_052503}, is based on a peculiar effect of radiation trapping in inhomogeneous magnetized plasma. Under specific conditions, oblique launch of gyrotron radiation results in generation of right-hand-polarized (R) electromagnetic waves propagating with high longitudinal refractive index $N_{\|}$ in the vicinity of the cyclotron resonance layer, which leads to effective single-pass absorption of the injected microwave power. The theory developed in the present paper is related to the final stage ECRH when the trapped electromagnetic wave approaches the resonant surface, where the wave frequency $\omega$ is equal to the local electron cyclotron frequency $\omega_{e}(\vec{r})$.

This paper is organized in the following way. To begin with, in Sec.~\ref{s2} we describe the magnetic field in a linear trap in the paraxial approximation. In Sec.~\ref{s3} we derive the dispersion equation for electromagnetic waves near the electron cyclotron frequency. In Sec.~\ref{s4}, we find parametric formulas for the ray trajectories and derive a criterion for the resonance point on the plasma axis to attract the ray trajectories.  In Sec.~\ref{s5}, the ray trajectories are analysed and systemized. Finally, in Sec.~\ref{s6}, the effect of finite plasma pressure on the electron cyclotron resonance heating is discussed. A short summary is given in Sec.~\ref{s7}.

%
%

\section{Magnetic field in paraxial approximation}
\label{s2}

Neglecting the the finite $\beta$ effects (where $\beta$ is the ratio of plasma pressure to the pressure of magnetic field), we assume that the magnetic field in the plasma is potential, that is
    \begin{equation}
    \label{2:1}
      \vec{B}=\nabla \psi
      ,
    \end{equation}
where the potential $\psi$ obeys the Laplace equation $\nabla^2\psi=0$. For an axisymmetric system in cylindrical system of coordinates the latter has the form
    \begin{equation}
    \label{2:2}
    \frac{1}{r}\frac{\partial}{\partial r}\left( r\frac{\partial \psi}{\partial r} \right)
    +
    \frac{\partial^2 \psi}{\partial z^2}=0
    .
    \end{equation}
Putting here
    \begin{equation}
    \label{2:3}
    \psi=\sum\limits_{n} A_{n}(z)r^n
    \end{equation}
yields a recursive formula
    \begin{equation*}
    A_{n}''+(n+2)^2A_{n+2}=0
    \end{equation*}
for the functions $A_{n}(z)$. It allows to express all the functions $A_{n}(z)$ either through $A_{0}(z)$ or $A_{1}(z)$. Since the radial component $B_r(r,z)$ of $\vec{B}$ is zero at $r=0$ due to axial symmetry, the coefficient $A_{1}(z)$ should be zero, and the  coefficient $A_{0}(z)$ can be expressed through the magnetic field $B_{0}(z)=B_z(0,z)$ on the system axis. Paraxial approximation means that high-$n$ terms in \eqref{2:3} are small and can be dropped. Keeping first two non-zero terms of the expansion, we obtain
    \begin{equation}
    \psi(\vec{r})=\int\limits^{z}dz B_{0}(z)-\frac{r^2}{4}B_{0}'(z)
    ,
    \end{equation}
where the prime stands for the derivative over $z$. Consequently,
    \begin{equation}
    B_{r}(\vec{r})
    =
    \parder{\psi}{r}
    =
    -\frac{r}{2}B_{0}'(z)
    ,
    \qquad
    B_{z}(\vec{r})
    =
    \parder{\psi}{z}
    =
    B_{0}(z)
    -
    \frac{r^2}{4}B_{0}''(z)
    .
    \end{equation}
Following Ref.~\cite{ZvonkovTimofeev1988SovJPlasmaPhys_14_743}, we expand the modulus of the magnetic field strength
    \begin{equation}
    \label{2:7}
    B(\vec{r})
    =
    \sqrt{B_{r}^2+B_{z}^2}
    \approx
    B_{0} (z)-\frac{r^2}{4}\left( B_{0}''-\frac{1}{2} \frac{B_{0}'^2}{B_{0}} \right)
    \end{equation}
about the point $z=z_s$ of the electron cyclotron resonance on the system axis to put it in the form
    \begin{equation}
    \label{2:8}
    B(\vec{r})
    =
    B_{0}(z_{s})
    \left(
        1 + \frac{z-z_{s}}{L_{\parallel}} - \frac{r^2}{L_{\perp}^2}
    \right)
    ,
    \end{equation}
where
    \begin{equation}
    \label{1.1}
    \frac{1}{L_{\parallel}}
    =
    \frac{B_{0}'(z_{s})}{B_{0}(z_{s})}
    ,\qquad
    \frac{1}{L_{\perp}^2}
    =
        \frac{1}{4}
        \frac{B_{0}''(z_{s})}{B_{0}(z_{s})}
        -
        \frac{1}{8}
        \frac{B_{0}'^{2}(z_{s})}{B_{0}^{2}(z_{s})}
    .
   \end{equation}

\begin{figure}
  \includegraphics[width=0.5\textwidth]{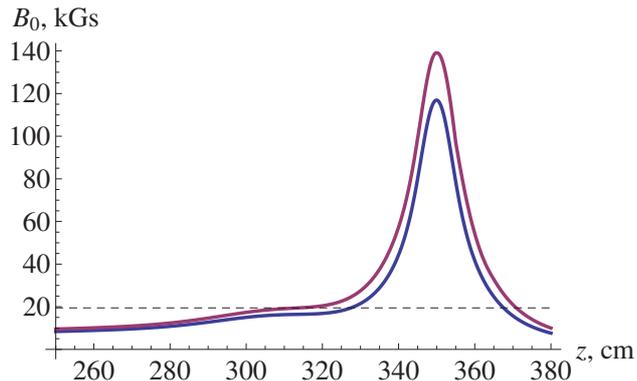}
  \caption{
    On-axis distribution of the magnetic field strength $B_0(z)$ in the GDT device for two configurations with maximum magnetic field $117\,\text{kGs}$ and $145\,\text{kGs}$; the dashed line shows the magnetic field strength corresponding to the electron cyclotron resonance at the frequency $54.5\,\text{GHz}$.
    }
  \label{figure:1}
\end{figure}

\begin{table}
\begin{tabular}{|c||c|c|c|c|c|c|} \hline
  $\omega/2\pi$ (GHz) &$B_{\max}$ (kGs)
    &$z_{s}$ (cm)  & $L_{\parallel}$ (cm) & $L_{\perp}^2$ (cm$^2$)
    & $L_{\|}^2/L_{\perp}^2$
  \\ \hline
  $54.5$    & $116.9$
    & $327.5$  &  $\phantom{0}29.8$   &  $\phantom{-0}770.7$
    & $\phantom{-}1.16$
  \\ \hline
  $54.5$    & $145.4$
    & $306.6$  &  $136.3$   &  $-3768.0$
    & $-4.93$
  \\ \hline
\end{tabular}
    \caption{
        The characteristics of the electron cyclotron resonances in GDT.
    }
    \label{tab:1}
\end{table}
As an example of general theory implementation, we consider ECRH in GDT at the frequency $\omega/2\pi=54.5\,\text{GHz}$. We consider two magnetic field configurations, which differ by the maximum of the field strength $B_{\max}$ at the magnetic mirrors. Low field configuration is characterized by $B_{\max}=117\,\text{kGs}$, and strong field configuration has $B_{\max}=145\,\text{kGs}$. Fig.~\ref{figure:1} shows plots of the magnetic field profile $B_0(z)$ in the west half of GDT for these two configurations, and the horizontal line indicate the resonant magnitude of the magnetic field $B_\text{res}=19.5\,\text{kGs}$.
Resonance points $z_s$ are located at the intersection of this line with the plots of $B_0(z)$. Computed values of the parameters $z_s$, $L_{\parallel}$, $L_{\perp}^2$ are given in Table \ref{tab:1}.

\section{Ray hamiltonian}
\label{s3}

To derive the ray hamiltonian, we shall use the helical (spiral) coordinates
    \begin{gather*}
    \dif{s}^{\pm}
    = \frac{\dif{x} \pm i \dif{y}}{\sqrt{2}}
    ,\quad
    \dif{s}^{\|}=\dif{z}
    .
    \end{gather*}
The metric tensor can be found from the identity
    \begin{gather*}
    \dif{s}^2
    =
    2\dif{s}^{+}\dif{s}^{-} + \dif{s}^{\|}\dif{s}^{\|}
    \equiv
    g_{ik}\dif{s}^{i}\dif{s}^{k}
    \end{gather*}
and has the form
    \begin{gather*}
    g_{ik}=
    \begin{pmatrix}
      0 & 1 & 0\\
      1 & 0 & 0\\
      0 & 0 & 1
    \end{pmatrix}
    ,
    \end{gather*}
which immediately yields the rule
    \(
    \dif{s}_{\pm}=\dif{s}^{\mp}
    \),
    \(
    \dif{s}_{\|}=\dif{s}^{\|}
    \)
for the covariant components
    \(
    \dif{s}_{i} = g_{ik} \dif{s}^{k}
    \).
Components of other vectors are defined similarly. In particular, helical components of the electric field $\vec{E}$ and normalized wave vector $\vec{N}=c\vec{k}/\omega$ (refractive index) are
    \begin{gather*}
    E^{\pm}
    =\frac{E_x \pm i E_y}{\sqrt{2}}
    =E_{\mp}
    ,\qquad
    E^{\|}
    =E_z
    =E_{\|}
    ,
    \\
    N^{\pm}
    =\frac{N_x \pm i N_y}{\sqrt{2}}
    =N_{\mp}
    ,\qquad
    N^{\|}
    =N_z
    =N_{\|}
    .
    \end{gather*}
Contravariant components of the permittivity tensor $\varepsilon^{\alpha\beta}$ are computed through those in the cartesian coordinates $\varepsilon_{ij}$ by the standard rules. For example
\begin{gather*}
    \varepsilon^{++}
    = \parder{s^{+}}{x^i}\parder{s^{+}}{x^j}\varepsilon_{ij}
    =  \frac{1}{2}\left(
        \varepsilon_{xx}-\varepsilon_{yy}+i\varepsilon_{xy}+i\varepsilon_{yx}
    \right)
    .
\end{gather*}
In the cold plasma limit, the diagonal elements $\varepsilon^{++}$ and $\varepsilon^{--}$ are zero whereas the off-diagonal components are not:
\begin{gather*}
    \varepsilon^{+-}
    = \parder{s^{+}}{x^i}\parder{s^{-}}{x^j}\varepsilon_{ij}
    =  \frac{1}{2}\left(
        \varepsilon_{xx}+\varepsilon_{yy}-i\varepsilon_{xy}+i\varepsilon_{yx}
    \right)
    \equiv \varepsilon_{-}
    ,
    \\
    \varepsilon^{-+}
    = \parder{s^{-}}{x^i}\parder{s^{+}}{x^j}\varepsilon_{ij}
    =  \frac{1}{2}\left(
        \varepsilon_{xx}+\varepsilon_{yy}+i\varepsilon_{xy}-i\varepsilon_{yx}
    \right)
    \equiv \varepsilon_{+}
    .
\end{gather*}
All other components except for $\varepsilon^{\|\|}=\varepsilon_{zz}\equiv \varepsilon_{\|}$ are also zero so that
\begin{gather*}
    \varepsilon^{\alpha\beta}
    =
    \begin{pmatrix}
      0 & \varepsilon_{+} & 0 \\
      \varepsilon_{-} & 0  & 0\\
      0 & 0 & \varepsilon_{\|}
    \end{pmatrix}
    ,
\end{gather*}
where
    \begin{gather}
    \label{3:17}
    \varepsilon_{\pm}
    = 1 - \sum_{s}\frac{\omega_{ps}^2}{\omega(\omega\mp\Omega_{s})}
    \approx
    1 - \frac{\omega_{pe}^2}{\omega(\omega\pm\omega_{e})}
    ,
    \qquad
    \varepsilon_{\|}
    = 1 - \sum_{s}\frac{\omega_{ps}^2}{\omega^2}
    \approx
    1 - \frac{\omega_{pe}^2}{\omega^2}
    ,
    \end{gather}
summation goes over the plasma species $s$, $\omega_{ps}=\sqrt{4\pi e_{s}^2n_{s}/m_{s}}$, $\Omega_s= e_{s}B/m_{s}c$, and $\omega_{e}=-\Omega_{e}=|e|B/m_{e}c$.  Mixed permittivity tensor is diagonal in the cold plasma. In particular,
\begin{gather*}
    \varepsilon^{\alpha}_{\cdot\beta}
    =
    \varepsilon^{\alpha\gamma}g_{\gamma\beta}
    =
    \begin{pmatrix}
      \varepsilon_{+} & 0 & 0 \\
      0 & \varepsilon_{-} & 0 \\
      0 & 0 & \varepsilon_{\|}
    \end{pmatrix}
    .
    \end{gather*}

In the approximation of geometric options, the wave equation reads
\begin{gather*}
    \left(
        \varepsilon^{\alpha}_{{}\cdot\beta}
        -
        N^2 g^{\alpha}_{{}\cdot\beta}
        +
        N^{\alpha}N_{\beta}
    \right)
    E^{\beta}
    =0
    .
    \end{gather*}
It leads to the dispersion equation \cite{Timofeev2009(en)}
    \begin{gather}
    \label{3:15}
    1
    + \frac{N^{+}N_{+}}{\varepsilon_{+}-N^2}
    + \frac{N^{-}N_{-}}{\varepsilon_{-}-N^2}
    + \frac{N^{\|}N_{\|}}{\varepsilon_{\|}-N^2}
    =0
    ,
    \end{gather}
and the polarization of the eigenmodes is described by the vector
    \begin{gather}
    \label{3:16}
    \vec{E}
    =
    \{
        E^{+},E^{-},E^{\|}
    \}
    \propto
    \biggl\{
            \frac{N^{+}}{\varepsilon_{+}-N^2}
        ,
            \frac{N^{-}}{\varepsilon_{-}-N^2}
        ,
        \frac{N^{\|}}{\varepsilon_{\|}-N^2}
    \biggr\}
    .
    \end{gather}
Note that $N^{+}N_{+}=N^{-}N_{-}=\tfrac{1}{2}(N_{x}^2+N_{y}^2)= \tfrac{1}{2}N^2_{\bot}$, and $E^{+}$ and $E^{-}$ are the amplitudes of the left and right circular wave components that rotate in the direction of the electron and ion cyclotron gyration, respectively.

Below we restrict ourselves to the case $\omega\simeq\omega_{e}$, where $\varepsilon_{-}\to\infty$.

Assuming that $N^2 \ll \varepsilon_{-}$, we can drop the third summand in Eq.~\eqref{3:15}. Then, for transverse propagation, $N_{\|}\to 0$, we find two eigenmodes.  The ordinary mode is polarized along the magnetic field:
    \begin{gather*}
    N^2\simeq \varepsilon_{\|}
    ,
    \qquad
    \vec{E} \simeq \{0, 0, 1\}
    .
    \end{gather*}
It is obtained from the dispersion equation \eqref{3:15} by evaluating the $0/0$ uncertainty in the last term. The extraordinary mode is circularly left polarized (and gyrates in the ion side):
    \begin{gather*}
    N^2\simeq 2\varepsilon_{+},
    \qquad
    \vec{E} \simeq \{1, 0, 0\}
    .
    \end{gather*}

In the case of longitudinal propagation, $\theta=N_{\bot}/N \to 0$, the ordinary node can still be found using the assumption $N^2\ll\varepsilon_{-}$. It has the left (ion) circular polarization:
    \begin{gather*}
    N^2
    \simeq
    \varepsilon_{+}
    \left[
        1
        +
        \frac{\varepsilon_{\|}-\varepsilon_{+}}{2\varepsilon_{\|}}
        \,
        \theta^2
    \right]
    ,
    \qquad
    \vec{E} \simeq \{1, 0, 0\}
    .
    \end{gather*}

Dispersion and polarization of the extraordinary wave undergo significant changes as the angle $\theta$ approaches zero. First of all it is worth noting that $N^2=\varepsilon_{-}$ is the exact solution at $\theta=0$ and, hence, $N^2\to\infty$ as $\varepsilon_{-}\to \infty$ so that the assumption $N^2\ll \varepsilon_{-}$ breaks.  Proceeding with greater accuracy as demonstrated in Ref.~\cite{ZvonkovTimofeev1988SovJPlasmaPhys_14_743}, we put
    \begin{gather*}
    \frac{N^{+}N_{-}}{\varepsilon_{+}-N^2}
    \simeq
    -\frac{N_{\bot}^{2}/2}{N^2}
    ,
    \qquad
    \frac{N^{\|}N_{\|}}{\varepsilon_{\|}-N^2}
    \simeq
    -1 + \frac{N_{\bot}^{2}-\varepsilon_{\|}}{N^2}
    \end{gather*}
into Eq.~\eqref{3:15} to obtain the reduced dispersion equation
    \begin{gather}
    \label{3:25}
    \varepsilon_{\parallel}N^2
    +
    \tfrac{1}{2}\varepsilon_{-} N_{\perp}^2
    -
    \varepsilon_{\|}\varepsilon_{-}
    =0
    .
    \end{gather}
Its solution
    \begin{gather*}
    N^2 =
    \tfrac{
        \displaystyle
        \varepsilon_{\|}\varepsilon_{-}
    }{
        \displaystyle
        \varepsilon_{\|}
        +
        \tfrac{1}{2}\theta^2 \varepsilon_{-}
    }
    \end{gather*}
shows that the dispersion law abruptly changes at $\theta\simeq \theta_{l} = \sqrt{2\varepsilon_{\|}/\varepsilon_{-}}$ \cite{Timofeev1992SovJPlasmaPhys_18_214, Timofeev2009(en)} since
$N^2 \simeq \varepsilon_{-}$ for $\theta\ll \theta_{l}$ whereas $N^2 \simeq 2 \varepsilon_{\|}/\theta^2$ and $N_{\bot}^2 \simeq 2 \varepsilon_{\|}$  for $\theta_{l} \ll \theta \ll \pi/2$.

In the range of frequencies $\omega>\omega_{e}$, where $\varepsilon_{-}<0$, a plasma resonance, characterized by $N\to\infty$, occurs at the critical angle $\theta_\text{cr}=\sqrt{-\varepsilon_{\|}/2\varepsilon_{-}}$ (we assume that $\varepsilon_{\|}>0$, i.e. $\omega>\omega_{pe}$), and the plasma is transparent only for $\theta>\theta_\text{cr}$. It means that a packet of extraordinary wave can hardly access the surface of electron cyclotron resonance, where $\omega=\omega_{e}$, from the side of low magnetic field. This fact will be confirmed by direct calculations in Sec.~\ref{s5}.

For $\theta \gg \theta_{l}$, the wave field has almost complete longitudinal polarization ($E_{\|} \gg E_{+} \gg E_{-}$) since
\begin{gather*}
    \vec{E}
    \propto
    \biggl\{
        -\frac{\varepsilon_{-}\theta^2}{2\varepsilon_{\|}} ,
        1 ,
        - \frac{\varepsilon_{-}\theta}{\sqrt{2}\varepsilon_{\|}}
    \biggr\}
    .
    \end{gather*}
In the range $\theta \ll \theta_{l}$, the polarization is given by the vector
    \begin{gather*}
    \vec{E}
    \propto
    \biggl\{
        -\frac{\varepsilon_{-}\theta^2}{\varepsilon_{\|}} ,
        1 ,
        - \frac{\varepsilon_{-}\theta}{\sqrt{2}\varepsilon_{\|}}
    \biggr\}
    ,
    \end{gather*}
and the left circular component is small ($E^{+}\ll E^{-}, E^{\|}$). The polarization drastically rearranges at the angle $\theta\simeq \theta_{l}^2$.
In the interval $\theta_{l}^2\ll\theta\ll\theta_{l}$, the field still has almost complete longitudinal polarization ($E_{\|} \gg E_{-} \gg E_{+}$) as in the case
$\theta\gg\theta_{l}$. On the contrary, for $\theta \ll \theta_{l}^2$, the wave field has almost complete right circular polarization that rotates in the direction of the electron gyration ($E_{-} \gg E_{\|} \gg E_{+}$).

It is important to note that $N_{\bot}$ remains finite in the limit $\varepsilon_{-}\to\infty$. This allows us to substitute the first term in Eq.~\eqref{3:25} with $\varepsilon_{\|}N_{\|}^2$ since $\varepsilon_{\|}N_{\bot}^2$ is small as compared to the second term. Second simplification is due the observation that near the electron cyclotron resonance, at $\omega\approx\omega_e$, it is sufficient to keep only the last term in the expression \eqref{3:17} for $\varepsilon_{-}$. Then, Eq.~\eqref{3:25} takes the form
    \begin{equation}
    \label{3:29}
    D
    =
    \varepsilon_{\parallel}N_{\parallel}^2
    \frac{\omega (\omega_{e}(\vec{r})-\omega)}{\omega_{pe}^2}
    +
    \frac{1}{2}N_{\perp}^2-\varepsilon_{\parallel}
    =0
    ,
    \end{equation}
where the factor
    \begin{equation}
    \label{3:30}
    (\omega_{e}(\vec{r})-\omega)
    =
    \omega
    \left(
        \frac{z-z_{s}}{L_{\parallel}} - \frac{r^2}{L_{\perp}^2}
    \right)
    \end{equation}
is computed using Eq.~\eqref{2:8}.

In a weakly inhomogeneous magnetic field,  Eq.~\eqref{3:29} defines slowly varying function $D$ of the radius-vector $\vec{r}$. Given the validity of geometric optics, the function $D=D(\vec{k},\vec{r}; \omega)$ can be considered as the ray Hamiltonian \cite{Stix1992} that governs motion of a wave packets similar to the ordinary Hamiltonian that governs motion of a particle in mechanics. Appropriate set of equation
    \begin{align}
    \label{3:34}
    \der{\vec{r}}{t}
     &=
    -
    \frac{\partial D}{\partial \vec{k}}
    ,
    &
    \der{\vec{k}}{t}
    &=
    \phantom{-}
    \frac{\partial D}{\partial \vec{r}}
    \end{align}
relates the radius-vector $\vec{r}$ and the wave-vector $\vec{k}$ of the packet with some ``time'' $t$ which, however, has neither implicit meaning nor dimension of physical time.

\begin{figure}
  \includegraphics[width=0.5\textwidth]{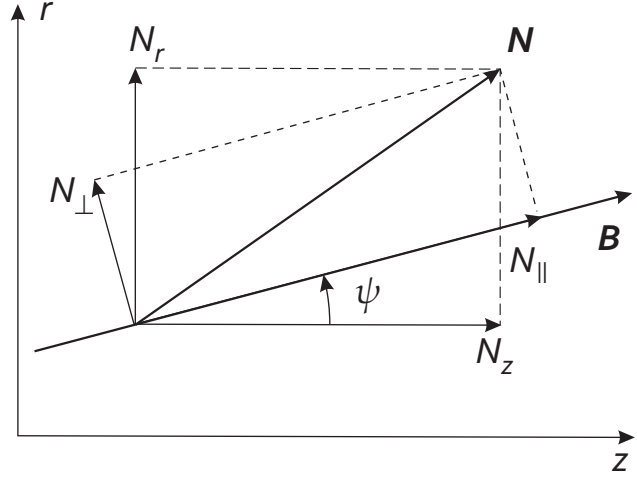}
  \caption{
    The effect of magnetic field lines inclination.
  }
  \label{figure:2}
\end{figure}
The transversal and longitudinal components, $N_{\bot}$ and $N_{\|}$, of the vector $\vec{N}$ can be matched to the radial and axial projections, $N_{r}$  and $N_{z}$, as shown in Fig.~\ref{figure:2}, where $\psi \approx B_r/B_0 = -r/2L_{\parallel}$ is the inclination angle of  the magnetic field lines:
    \begin{equation*}
    N_{\perp}=\cos\psi N_r-\sin\psi N_{z}
    ,
    \qquad
    N_{\parallel}=\cos\psi N_z+\sin\psi N_{r}
    .
    \end{equation*}
Since $\psi\ll 1$ and $N_{\bot}\ll N_{\|}$ we have approximately
    \begin{equation}
    \label{3:32}
    N_{\perp} \approx N_r-\psi N_{z} \qquad
    N_{\parallel} \approx N_z
    .
    \end{equation}

To proceed further, we introduce a dimensionless parameter
    \[
    \alpha=\frac{\omega_{pe}^2}{4\varepsilon_{\parallel}\omega^2}
    \]
and normalize coordinates on the parameter of length
    \[
    \ell=2\alpha L_{\parallel}=\frac{\omega_{pe}^2L_{\parallel}}{2\varepsilon_{\parallel}\omega^2}
    \]
so that
    \[
    \rho=r/\ell
    ,
    \qquad
    \chi = (z-z_{s})/\ell
    .
    \]
Instead of the dimensionless vector $\vec{N}$, we introduce the dimensionless vector of the wave momentum $\vec{p}=(\omega\ell/c)\vec{N} $ with the components
    \[
    p_{\rho}=\frac{\omega \ell N_{r}}{c}
    ,
    \qquad
    p_{\chi}=\frac{\omega \ell N_{z}}{c}
    .
    \]
With all these amendments being combined, the Hamiltonian \eqref{3:29} takes the form
    \begin{equation}
    \label{3:37}
    D=(\chi -\mu \rho^2)p_{\chi}^2
    +(p_{\rho}+\alpha \rho p_{\chi})^2
    -\nu
    =0
    ,
    \end{equation}
where
    \begin{gather*}
    \mu=2\alpha \left(\frac{L_{\parallel}}{L_{\perp}} \right)^2
    = \frac{1}{2\varepsilon_{\parallel}}\left(\frac{\omega_{pe}}{\omega}\right)^2 \left(\frac{L_{\parallel}}{L_{\perp}}\right)^2 ,
    \\
    \nu=2 \alpha   \left(\frac{\omega_{pe} L_{\parallel}}{c}\right)^2
    =
    \frac{\omega_{pe}^4}{\omega^2 c^2}
    \frac{L_{\parallel}^2}{2\varepsilon_{\parallel}}
    .
    \end{gather*}
In Ref.~\cite{ZvonkovTimofeev1988SovJPlasmaPhys_14_743} the term term $\alpha\rho p_{\chi}$ in Eq.~\eqref{3:37} was lost since $N_{\bot}$ was set equal to $N_{r}$.
In what follows we assume that $\alpha>0$, i.e. $\omega>\omega_{pe}$. ECR heating is not effective in the case $\omega<\omega_{pe}$ as will be discussed later.

\section{Stability of the resonance zone}
\label{s4}

Following to Ref.~\cite{ZvonkovTimofeev1988SovJPlasmaPhys_14_743}, we consider only those ray trajectories that lay in a plane passing through the axis $z$ of the system. Appropriate ray trajectories for the Hamiltonian \eqref{3:37} are governed by the equations
\begin{align}
  \label{4:1}
  \der{\rho}{\tau}
  &=
  \phantom{-}
  \frac{\partial D}{\partial p_{\rho}}  = 2(p_{\rho}+\alpha \rho p_{\chi}),
  \\
  \label{4:2}
  \der{p_{\rho}}{\tau}
  &=
  -\frac{\partial D}{\partial \rho}  = 2\mu \rho p_{\chi}^2-2\alpha p_{\chi} (p_{\rho}+\alpha \rho p_{\chi}),
  \\
  \label{4:3}
  \der{\chi}{\tau}
  &=
  \phantom{-}
  \frac{\partial D}{\partial p_{\chi}}
  =
  2(\chi-\mu\rho^2)p_{\chi}
  +2\alpha\rho (p_{\rho}+\alpha\rho p_{\chi})
  ,
  \\
  \label{4:4}
  \der{p_{\chi}}{\tau}
  &=
  -\frac{\partial D}{\partial \chi}=-p_{\chi}^2
  ,
\end{align}
where $\tau$ is the new dimensionless time. The last equation is separated from others and has the solution $p_{\chi}=1/(\tau - C_1)$, which depend on the arbitrary constant $C_1$. The latter can be set to zero by appropriate choice of the origin of time; hence
    \begin{equation}
    \label{4:5}
    p_{\chi}=1/\tau
    .
    \end{equation}
Putting \eqref{4:5} into Eqs.~\eqref{4:1} and~\eqref{4:2}, we note that their solution on the interval $\tau>0$ can be sought in the form
    \begin{gather}
    \label{4:6}
    \rho=A_{+} \tau^{\gamma_{+}}+A_{-} \tau^{\gamma_{-}}
    ,
    \\
    \label{4:7}
    p_{\rho}=B_{+}\tau^{\gamma_{+}-1}+B_{-}\tau^{\gamma_{-}-1}
    .
    \end{gather}
It is then easy to find that
    \begin{gather}
    \label{4:8}
    \gamma_{\pm}
    =
    \frac{1}{2}
    \left(
        1\pm \sqrt{1-8\alpha+16\mu}
    \right)
    ,
    \\
    B_{\pm}
    =
    \frac{1}{2}
    \left(
        \gamma_{\pm}-2\alpha
    \right)
    A_{\pm}
      .
    \end{gather}
The function $\chi$ can be found from Eq.~\eqref{3:37}:
    \begin{multline}
    \label{4:9}
    \chi=\mu \rho^2+\nu \tau^2-\left(\tau p_{\rho}+\alpha\rho\right)^2 =
    \\
    \nu \tau^2
    +
    (4 \mu-\alpha )
    A_{-} A_{+} \tau
    -
    \frac{1}{4}
    \left(2\gamma_{+}-4 \alpha\right)
    A_{+}^2 \tau ^{2\gamma_{+}}
    -
    \frac{1}{4}
    \left(2\gamma_{-}-4 \alpha\right)
    A_{-}^2 \tau ^{2\gamma_{-}}
    .
    \end{multline}
The obtained solution is valid for $\rho\ll 1$, $\chi\ll 1$, since it is based on the paraxial approximation \eqref{2:8}. Second condition of validity reads $p_{\rho}/p_{\chi}=p_{\rho}\tau\ll 1$; it guaranties that $\theta=p_{\rho}\tau+\alpha\rho \ll 1$. In terms of the time $\tau$ the validity region is limited by the inequalities $|A_{+}\tau^{\gamma_{+}}|\ll 1$ and $|A_{-}\tau^{\gamma_{-}}|\ll 1$.

The functions $\rho(\tau)$ and $\chi(\tau)$ determine a ray trajectory in the parametric form. They depend on the two amplitudes $A_{+}$ and $A_{-}$, which can be related to the initial coordinates of the ray $\rho_{0}$ and $\chi_{0}$ at a given initial value of the parameter $\tau_{0}$. We assume that a particular trajectory starts at some $\tau_{0}>0$ and ends at $\tau=0$. Note that ECRH occurs at $\tau\to 0$ when $p_{\chi}\to \infty$ (i.e. $N_{\|}\to\infty$) and the extraordinary waves transforms to small-scale oscillations that are effectively absorbed by plasma in single pass of the injected wave.

Absorption of the wave energy on intermediate parts of the ray trajectory is weak even if the trajectory  crosses the resonance surface, where $\omega=\omega_{e}$; the latter is described by the equation
    \begin{equation}
    \label{4:9'}
    \chi = \mu\rho^2
    \end{equation}
in dimensionless notations.

It is readily seen that both $\rho$ and $\chi$ tend to zero at $\tau\to0$, if the real parts of both $\gamma_{+}$ and $\gamma_{-}$ are positive, $\re \gamma_{\pm}>0$. This occurs if and only if $1-8\alpha+16\mu<1$ and means that any ray trajectory reaches the resonance point $\rho=\chi=0$ on the axis of the system. In this case the point $\rho=\chi=0$ turns out to be an attractor for the ray trajectories. In the opposite case the resonance point repels most of the trajectories (see below).

\begin{figure}
  \includegraphics[width=0.5\textwidth]{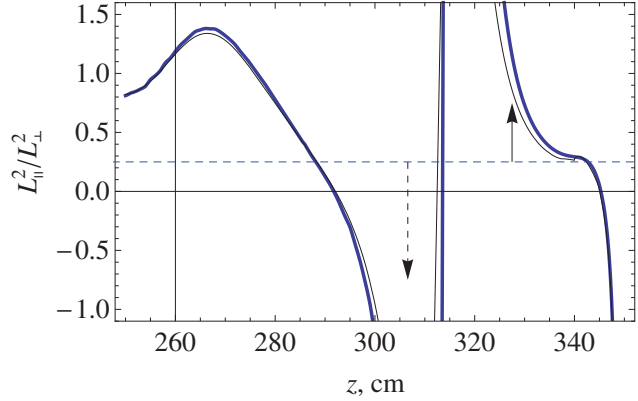}
  \caption{(Color online)
    Profile of the ratio $L_{\|}^2/L_{\bot}^2$ in the GDT device for the low (solid line) and strong (thin line) magnetic field configurations. Attractor region is located below the dashed line. Vertical solid and dashed arrow show position of the fundamental cyclotron resonance at the frequency $54.5\,\text{HGz}$ for the low and strong field configurations, respectively. The low-field-resonance falls into the repeller region, and the strong-field-resonance is located within the attractor region.
  }
  \label{figure:3}
\end{figure}
For $\alpha>0$ the condition $1-8\alpha+16\mu<1$ is equivalent to the inequality
    \begin{equation}
    \label{4:10}
    \frac{L_{\parallel}^2}{L_{\perp}^2}
    <
    \frac{1}{4}.
    \end{equation}
The criterion \eqref{4:10} lays in-between those obtained in Ref.~\cite{ZvonkovTimofeev1988SovJPlasmaPhys_14_743}, where it was equivalent to ${L_{\parallel}^2}{L_{\perp}^2}<0$, and in Ref.~\cite{Gospodchikov+2011PPR_37_768}, where the inequality ${L_{\parallel}^2}{L_{\perp}^2}<\frac{1}{2}$ was derived from approximate reasonings.

As an example, Fig.~\ref{figure:3} plots the ratio ${L_{\parallel}^2}/{L_{\perp}^2}$ for the two magnetic field configurations in GDT referred to in table~\ref{tab:1}. Attractor region \eqref{4:10} is located below the dashed line in Fig.~\ref{figure:3}, and the resonances are indicated by vertical arrows. It is remarkable that the two plots (thick and thin curves for the low and strong field configurations, respectively) almost coincide whereas the positions of the resonant point $z_{s}$ differ significantly for the two configurations. Solid arrow stands for the low magnetic field configuration and indicates the resonance point $z_{s}=327.5\,\text{cm}$ located in the repeller region. Dashed arrow indicates the position of the resonant point $z_{s}=306.6\,\text{cm}$ for the strong magnetic field configuration; it is located in the attractor region.

Despite the fact that the resonance point on the axis for the low field configuration  does not satisfy the condition \eqref{4:10}, it can be reached by appropriately tuned ray trajectories as is shown in the next section.

\section{Ray trajectories}
\label{s5}


Depending on the parameter $\zeta = 1-8 \alpha +16\mu $, which is included in the definition \eqref{4:8} of the exponents $\gamma_{\pm}$, ray trajectories can be divided into three types, considered sequentially in the following three subsections.

\subsection{The case $0 < \zeta < 1$}
\label{ss5.1}

If $0 < \zeta < 1$, the exponents $\gamma_{\pm}$ are real and obey the inequalities $0<\gamma_{-}<\gamma_{+}<1$. It means that the on-axis resonant point $\rho=\chi=0$ serves as attractor for the ray trajectories.

\begin{figure}\centering
  \includegraphics[width=0.5\textwidth]{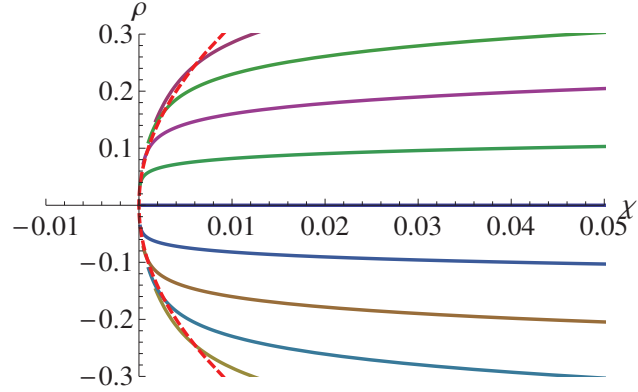}
  \caption{(Color online)
    Ray trajectories for $A_{+}=0$, $0<\zeta<1$, $\alpha=0.3$, $\mu=0.1$, $\nu=30$, $\gamma_{+}=0.72$, $\gamma_{-}=0.28$. Dashed line  indicates the resonance surface $\omega=\omega_{e}$. Ray trajectories approach the resonant point $\rho=\chi=0$ from the side of strong magnetic field.
  }
  \label{figure:4}
\end{figure}
To begin with, we consider a special case $A_{+}=0$. Then, it is possible to exclude the parameter $\tau$ and express the function $\chi$ through $\rho$. The result has the form
    \begin{equation}
    \label{5.1:1}
    \chi
    =
    \tilde{\nu} \rho^{2/\gamma_{-}}
    -
    \frac{1}{4}\left(\gamma_{-}
    -
    2\alpha \right) \rho^2
    ,
    \end{equation}
where $\tilde{\nu}=\nu/ A_{-}^{2/\gamma_{-}}$. The parameter $\nu$ is usually big and positive so that a typical trajectory approaches the resonant point at the axis from the side of strong magnetic field as shown in Fig.~\ref{figure:4}. This observation is in agreement with the fact, noted in Sec. \ref{s3}, that low magnetic field side is not transparent for the extraordinary waves.

\begin{figure}\centering
  \includegraphics[width=0.5\textwidth]{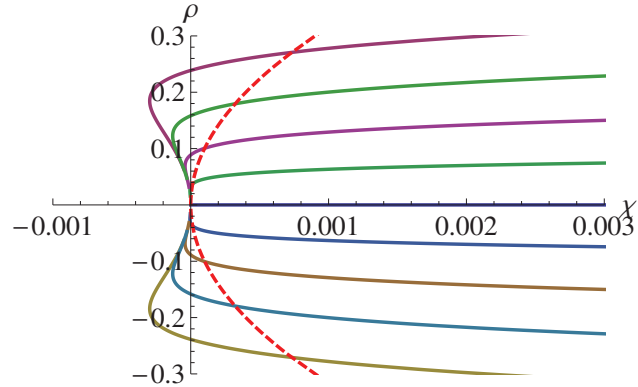}
  \caption{(Color online)
    Ray trajectories for $A_{+}=0$, $0<\zeta<1$, $\alpha=0.125$, $\mu=0.01$, $\nu=10$, $\gamma_{+}=0.7$, $\gamma_{-}=0.3$. Dashed line  indicates the resonance surface $\omega=\omega_{e}$. Ray trajectories approach the resonant point $\rho=\chi=0$ from the side of low magnetic field but starts on the strong field side.
  }
  \label{figure:5}
\end{figure}

Nevertheless there is a range of parameters, where the trajectory passes on the low field side just before accessing the resonant point as shown in Fig.~\ref{figure:5}. Indeed, since $\gamma_{-}<1$ the second term in Eq.~\eqref{5.1:1} dominates in the limit $\rho \to 0$, where the trajectory is described by the parabola
    \begin{equation}
    \label{5.1:1}
    \chi=-\frac{1}{8}\left(1- 4\alpha - \sqrt{1-8\alpha+16\mu}\right) \rho^2
    .
    \end{equation}
Its branches are directed to the side of positive $\chi$ except for the narrow region of the parameters $\alpha/2-1/16<\mu<\alpha^2$ that exists if $\alpha<\frac{1}{4}$. In this narrow region the ray trajectories approach the resonant point $\rho=\chi=0$ at the axis from the side of negative $\chi$ (i.e. low magnetic field) as shown in Fig.~\ref{figure:5}.

If both coefficients $A_{+}$ and $A_{-}$ are not zeros, qualitative picture remains the same and all the ray trajectories approach the resonant point $\rho=\chi=0$ as they do in 
Figs.~\ref{figure:4} and~\ref{figure:5}.
It means that wave packages deposit their energy mainly in the plasma core.

\subsection{The case $\zeta > 1$}
\label{ss5.2}

\begin{figure}\centering
  \includegraphics[width=0.5\textwidth]{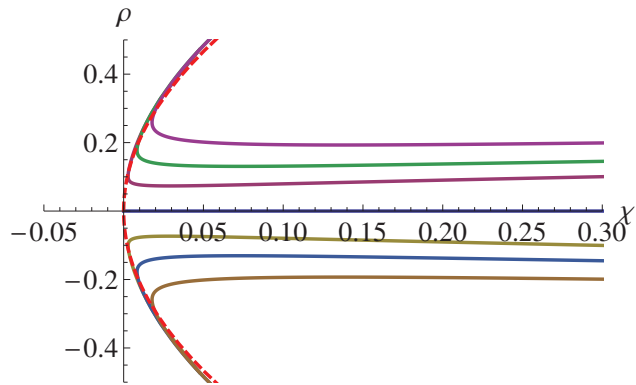}
  \caption{(Color online)
    Ray trajectories for $\zeta>1$, $A_{-} \neq 0$, $\alpha=0.5$, $\mu=0.9$, $\nu=5$, $\gamma_{+}=1.17$, $\gamma_{-}=-0.17$. Dashed line  indicates the resonance surface $\omega=\omega_{e}$.
  }
  \label{figure:6}
\end{figure}
If $\zeta>1$, both $\gamma_{+}$ and $\gamma_{-}$ are real but now
$\gamma_{+}>1$ whereas $\gamma_{-}<0$ and the term $A_{-}\tau^{\gamma_{-}}$ in Eq.~\eqref{4:6} dominates at $\tau \to 0$ except for the case where $A_{-}=0$. In other words, all trajectories with non-zero coefficient $A_{-}$ are repelled from the resonant point $\rho=\chi=0$ as shown in Fig.~\ref{figure:6} so that the resonant point cannot be reached by most of the ray trajectories.
%
\begin{figure}\centering
  \includegraphics[width=0.5\textwidth]{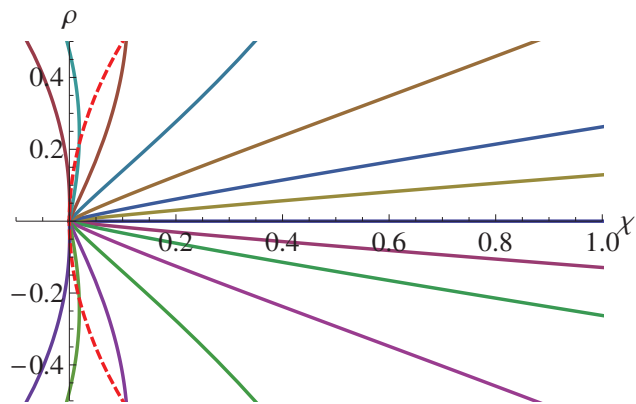}
  \caption{(Color online)
    Ray trajectories for $\zeta>1$, $A_{-}=0$,
    $\alpha=0.1$, $\mu=0.8$, $\nu=10$, $\gamma_{+}=2.30$, $\gamma_{-}=-1.30$.
    Dashed line  indicates the resonance surface $\omega=\omega_{e}$.
  }
  \label{figure:7}
\end{figure}
Nevertheless, special ray trajectories with $A_{-}=0$ can still reach the on-axis resonant point and deposit there the wave energy as shown in Fig.~\ref{figure:7}. These trajectories are described by the equation
    \begin{equation}
    \label{5.2:1}
    \chi=\hat{\nu}\rho^{2/\gamma_{+}}-\frac{1}{4}\rho^2 \left(\gamma_{+}- 2\alpha \right)
    ,
    \end{equation}
where $\hat{\nu}=\nu/A_{+}^{2/\gamma_{+}}$. In the limit $\rho \to 0$, the first term in Eq.~\eqref{5.2:1} dominates. Since it is positive, these special trajectories always approach the resonant point from the side of strong magnetic field.

Another point to note for the case $\zeta>1$ is where the repelled trajectories deposit the wave energy. As it was noted in Sec. \ref{s4}, the energy is deposited into plasma at $\tau\to 0$ when $p_{\chi}\to\infty$. In the case $A_{-}\neq 0$, the radial coordinate $\rho$ of a repelled trajectory formally tends to infinity at $\tau\to 0$. It might occur that $p_{\chi}$ will reach sufficiently large values before the ray will leave the plasma and some part of the wave energy will be deposited at the plasma periphery. As can be seen from Fig.~\ref{figure:6}, in such a case the wave energy is deposited in the vicinity of the resonant surface \eqref{4:9'}.


The case of $\zeta > 1$ is realized in GDT for the low magnetic field configuration, where $\alpha$ varies from $0.02$ to $0.5$, $\mu$ from $0.015$ to $1.1$, and $\zeta$ from $1.1$ to $15$, assuming that the plasma density in the heating zone varies from $0.5\times 10^{13}$ to $2.5\times 10^{13}\,\text{cm}^{-3}$.

\subsection{The case $\zeta < 0$}
\label{ss5.3}

In the case $\zeta<0$, the exponents $\gamma_{+}$ and $\gamma_{-}$ are complex conjugate to each other. The coefficients $A_{+}$ and $A_{-}$ should also compose a complex conjugate pair in order for $\rho(\tau)$ to be real function. The latter can be rewritten as
    \begin{equation}
    \label{5.3:1}
	\rho(\tau) =
    A_1 \tau^{1/2} \cos\left(\tfrac{1}{2} \sqrt{1-8 \alpha+16 \mu}\ln\tau\right)
    +
    A_2 \tau^{1/2} \sin\left(\tfrac{1}{2}\sqrt{1-8 \alpha+16 \mu}\ln\tau\right)
    ,
    \end{equation}
where $A_{1}$ and $A_{2}$ are the new amplitudes. The functions $\chi(\tau)$ and $p_{\rho}(\tau)$ can then be found from Eqs.~\eqref{3:37} and~\eqref{4:1}, respectively.  They describe the oscillating trajectory, which ends in the resonant point on the plasma  axis. Consequently, this point turns out to be attractor for the ray trajectories as in the case $0<\zeta<1$, and the wave energy is mainly deposited near the attractor.

\begin{figure}\center
  \includegraphics[width=0.5\textwidth]{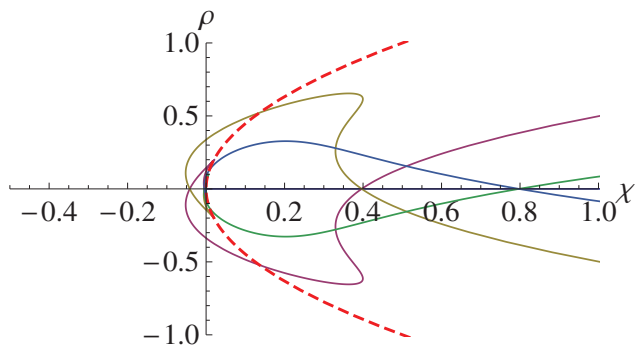}
  \caption{(Color online)
    Ray trajectories for $\zeta<0$, $\alpha=2$, $\mu=0.5$, $\nu=10$, $\re\gamma_{\pm}=0.5$, $\im\gamma_{\pm}= \pm 1.32$.
    Dashed line  indicates the resonance surface $\omega=\omega_{e}$.
  }
  \label{figure:8}
\end{figure}
Fig.~\ref {figure:8} shows an example of such trajectories. They can approach the resonance point mainly from the side of strong magnetic field while some trajectories transfer on the weak magnetic field side just before entering the on-axis resonance point.

In the GDT facility, the case $\zeta<0$ takes place for the strong field configuration as can be seen from table \ref{tab:1}. Numerical analysis of the ray trajectories for the strong field configuration performed in Ref.~\cite{Shalashov+2012PoP_19_052503} confirms the above conclusion that this configuration provides very effective ECRH.

\subsection{Accessibility of the on-axis resonance}
\label{ss5.4}

Suppose that
    \(
        \zeta < 1
    \)
so that an arbitrary ray trajectory ends near the on-axis resonance point $\rho=\chi=0$. It does not mean however that the point can be accessed from anywhere. Indeed, if we take an arbitrary point $(\rho_0,\chi_0)$ and arbitrary time $\tau_{0}$, we conclude from Eq.~\eqref{3:37} that they are related by the equation
    \begin{gather}
    \label{5.4:2}
    \chi_{0}
    =
    \mu \rho_{0}^2+\nu \tau_{0}^2-\left(\tau_{0} p_{\rho 0}+ \alpha\rho_{0}\right)^2
    ,
    \end{gather}
where
    \(
    \theta_{0}\equiv \tau_{0} p_{\rho 0} + \alpha \rho_{0}
    \)
stands for initial value of the angle $\theta\approx N_{\bot}/N$, which should be real for propagating waves as well as $\rho_{0}$ and $\chi_{0}$. Assuming that
$\alpha>0$ and, hence, $\nu>0$, we conclude from Eq.~\eqref{5.4:2} that for any
    \begin{gather}
    \label{5.4:4}
    \chi_{0}
    >
    \mu \rho_{0}^2
    -
    \theta_{0}^2
    \end{gather}
there exist such a time $\tau_{0}$ and a ray trajectory starting from a given point $(\rho_{0}, \chi_{0})$ at a given angle $\theta_{0}$, which reach the resonance point. Note that a trajectory with the angle $-\theta_{0}$ of opposite sign also reaches the resonant point from the same start point $(\rho_0,\chi_0)$ for the same time $\tau_0{}$.

The inequality \eqref{5.4:4} implies that a trajectory can for sure access the resonant point if it starts on the strong field side $\chi>\mu\rho^2$ of the resonant surface $\chi=\mu\rho^2$. As it is explained in Sec.~\ref{s3}, the low field side is not transparent for the extraordinary waves if $\alpha>0$ (i.e. $\omega>\omega_{pe}$).

In overdense plasma, where  $\alpha<0$ ($\omega<\omega_{pe}$), the parameter $\nu$ becomes negative, and instead of \eqref{5.4:4} we formally obtain the opposite inequality
    \(
    \chi_{0}
    <
    \mu \rho_{0}^2
    -
    \theta_{0}^2
    ,
    \)
which means that strong field side becomes opaque for the extraordinary waves, and the resonance point can be accessed mostly from the low field side. However the plasma periphery (where the plasma density is low and $\omega>\omega_{pe}$) will be opaque for the extraordinary waves in this case so that effective ECRH of overdense plasma is not possible as it is well known (see, e.g. \cite{Shalashov+2012PoP_19_052503}).

\section{Effect of finite $\beta$}
\label{s6}

So far we have ignored the effects that appear due to distortion of the magnetic field by the plasma pressure. Let now show that these effects are not important.

Equilibrium of plasma cylinder assumes that the sum of plasma pressure and the pressure of the magnetic field is constant across the plasma, i.e.
    \begin{equation}
    \label{6:1}
    p + \frac{B^2}{8 \pi} = \frac{B_0^2}{8 \pi}
    .
    \end{equation}
It follows from Eq.~\eqref{6:1} that small plasma pressure changes the magnetic field inside the plasma by small amount
    \begin{equation}
    \label{6:2}
    \delta B
    =
    -\frac{4 \pi p}{B_0}
    .
    \end{equation}
Assuming that the plasma pressure has parabolic radial profile,
    \begin{equation*}
    p
    =
    p_{0}
    \left(
        1-\frac{r^2}{a^2}
    \right)
    \end{equation*}
we conclude from Eq.~\eqref{6:2} that the parameter $L_{\bot}^2$ should be substituted with
    \begin{equation}
    \label{6:4}
    \frac{1}{L_{\perp}'^2}
    =
    \frac{1}{L_{\perp}^2}-\frac{\beta}{2a^2}
    ,
    \end{equation}
where $\beta={8 \pi p_{0}}/{B_{0}^2}$, and $a$ denotes the plasma radius in the heating zone. As a result, the criterion \eqref{4:10} takes the form
    \begin{equation}
    \label{6:5}
    \frac{L_{\parallel}^2}{L_{\perp}'^2}
    =
    \frac{L_{\parallel}^2}{L_{\perp}^2}\left(
        1- \frac{\beta}{2} \frac{L_{\perp}^2}{a^2}
    \right)
    <
    \frac{1}{4}
    .
    \end{equation}
It is seen from Eq.~\eqref{6:5} that the effect of plasma pressure is negligible if
${\beta}{L_{\perp}^2}/{2a^2}\ll 1$. In GDT, the factor ${\beta}{L_{\perp}^2}/{2a^2}$ does not exceed $0.01$ and, hence, the effect of plasma pressure can indeed be neglected.

\section{Conclusion}
\label{s7}



In this paper we have refined the theory initially developed in Ref.~\cite{ZvonkovTimofeev1988SovJPlasmaPhys_14_743} in order include the effect of magnetic field lines inclination as suggested in Ref.~\cite{Gospodchikov+2011PPR_37_768}. We have found analytic form for the parametric presentation of the ray trajectories near the axis of axially-symmetric linear trap and derived the criterion \eqref{4:10} that guaranties for all the ray trajectories to reach the point of electron cyclotron resonant on the trap axis. Our criterion \eqref{4:10} lays in-between those obtained in Ref.~\cite{ZvonkovTimofeev1988SovJPlasmaPhys_14_743}, where the effect of of magnetic field lines inclination was missed, and in Ref.~\cite{Gospodchikov+2011PPR_37_768}, where similar criterion was derived from approximate reasonings.



We have found that topology of the ray trajectories qualitatively changes as the  inhomogeneity of both the magnetic field strength and the field direction varies, which  significantly affects the efficiency of electron cyclotron heating of a plasma in a linear magnetic trap. In a case where the criterion \eqref{4:10} is satisfied, all the ray trajectories end near the on-axis resonant point which turns out to be attractor. In the opposite case, the on-axis resonant point repels the most of the ray trajectories but, still, there is a family of the ray trajectories that ends at that point, which guaranties effective ECRH.

As an example, we applied our theory to forthcoming ECRH experiment on the GDT facility \cite{Shalashov+2012PoP_19_052503}. Our theory leads to the conclusion that the efficiency of ECRH is very sensitive to the position of the on-axis resonant point which should be located within the attractor zone in order for the plasma interior to be heated. We have shown that the strong magnetic field configuration which was numerically analyzed in Ref.~\cite{Shalashov+2012PoP_19_052503} meets this requirements.

\begin{acknowledgments}

    The authors wish to thank P. A. Bagryansky for helpful discussions.
    This work has been supported by the Ministry of Education and Science of the Russian Federation (Grant No. 11.G34.31.0033), Presidium of the Russian Academy of Sciences (Program No. 30, 2012-2014).

\end{acknowledgments}



\begin{thebibliography}{1}

\bibitem{ZvonkovTimofeev1988SovJPlasmaPhys_14_743}
A.~V. Zvonkov and A.~V. Timofeev.
\newblock {\em Sov. J. Plasma Phys.}, 14:743, 1988.

\bibitem{Gospodchikov+2011PPR_37_768}
E.~Gospodchikov and O.~Smolyakova.
\newblock Ray trajectories near the electron cyclotron resonance surface in an
  axisymmetric magnetic trap.
\newblock {\em Plasma Physics Reports}, 37:768--774, 2011.
\newblock 10.1134/S1063780X11080034.

\bibitem{Shalashov+2012PoP_19_052503}
A.~G. Shalashov, E.~D. Gospodchikov, O.~B. Smolyakova, P.~A. Bagryansky, V.~I.
  Malygin, and M.~Thumm.
\newblock Auxiliary ecr heating system for the gas dynamic trap.
\newblock {\em Phys. Plasmas}, 19:052503, april 2012.

\bibitem{Ginzburg1970}
V.~L. Ginzburg.
\newblock {\em The propagation of electromagnetic waves in plasmas}.
\newblock Pergamon, Oxford, 2nd rev. and enl. edition, 1970.

\bibitem{Stix1992}
T.~H. Stix.
\newblock {\em Waves in Plasmas}.
\newblock AIP, New York, 1992.

\bibitem{Geller1996}
R.~Geller.
\newblock {\em Electron Cyclotron Resonance Ion Sources and ECR Plasmas}.
\newblock Institute of Physics Publishing, London, 1996.

\bibitem{Timofeev2009(en)}
A.~B. Timofeev.
\newblock {\em Resonance phenomena in the plasma oscillations}.
\newblock fizmatlit, Moscow, 2nd edition, 2009.
\newblock (in Russian).

\bibitem{PiliaFedorov1992}
A.~D. Piliya and V.~I. Fedorov.
\newblock {\em Linear conversion of electromagnetic and plasma waves in a two
  dimensionally inhomogeneous plasma}, pages 305--338.
\newblock American Institute of Physics, New York, 1992.

\bibitem{Timofeev1992SovJPlasmaPhys_18_214}
A.~V. Timofeev.
\newblock {\em Sov. J. Plasma Phys.}, 18:214, 1992.

\end{thebibliography}

%
\end{document}